\documentstyle[11pt,newpasp,twoside,epsf]{article}
\def\edcomment#1{\iffalse\marginpar{\raggedright\sl#1\/}\else\relax\fi}
\reversemarginpar

\begin{document}
\title{Galactic Superwinds Circa 2001}
\author{Timothy M. Heckman}
\affil{Department of Physics \& Astronomy, The Johns Hopkins University,
Baltimore, MD 21218}


\begin{abstract}
In this contribution I summarize our current knowledge of the nature
and significance of starburst-driven galactic winds (``superwinds'').
Superwinds are complex multiphase
outflows of cool, warm, and hot gas, dust, and magnetized
relativistic plasma.
The observational manifestations of superwinds result from the
hydrodynamical interaction between the primary energy-carrying wind fluid
and the ambient interstellar medium. 
Superwinds are ubiquitous
in galaxies in which the global star-formation rate per unit area
exceeds roughly 10$^{-1}$ M$_{\odot}$ yr$^{-1}$ kpc$^{-2}$. This
criterion is met by local starbursts and the high-z Lyman Break galaxies.
Several independent datasets and techniques imply that the total
mass and energy outflow rates
in a superwind are comparable to the starburst's
star-formation-rate and mechanical energy injection rate, respectively.
Outflow speeds in interstellar
matter entrained in the wind range from
$\sim 10^2$ to $10^3$ km/s, but the primary wind fluid
itself may reach velocities as high as $\sim$3000 km s$^{-1}$. The available
X-ray and far-UV ($FUSE$) data
imply that radiative losses in superwinds are not significant.
Superwinds may have established the mass-metallicity relation
in ellipticals and bulges, polluted the present-day inter-galactic
medium to a metallicity of $\sim$ 10 to 30\% solar, heated the
inter-galactic medium, and ejected
enough dust into the inter-galactic medium to have potentially observable
consequences.
\end{abstract}

\section{Introduction}

My interest in galactic winds was stimulated during the time I spent
at Steward Observatory as a Bart Bok Fellow during the early 1980's.
At that time, Ray Weymann was working on a theoretical model for
cosmic-ray-driven thermal winds in QSOs
(see Weymann et al. 1982), and I recall
discussing with Ray about how one might go about detecting the hydrodynamical
consequences of such flows on the surrounding interstellar and
intergalactic media.

These musings did not bear fruit until several years later
(McCarthy, van Breugel, \& Heckman 1987; Heckman, Armus, \& Miley 1987,
1990) when
my colleagues and I drew attention to evidence for global outflows
from the class of powerful, dusty
starbursts that had been discovered by IRAS.
By now, it is well-established that galactic-scale outflows of gas
(``superwinds'') are commonplace in the most actively star-
forming galaxies in both the local universe (e.g. Lehnert \& Heckman 1996;
Dahlem, Weaver, \& Heckman 1998; Veilleux et al. 1998) and
at high redshift (e.g. Pettini et al. 2001).

In this contribution, I will review 
the dynamical evolution of superwinds (section 2),
the nature and origin of their emission and absorption
(section 3),
their demographics (section 4), their estimated
outflow rates (section 5), and their likely fate (section 5).
Finally, I will describe their potential
implications for the evolution of galaxies and the inter-galactic
medium (section 6).

\section{The Conceptual Framework}

The engine that drives the observed outflows in
starbursts is the mechanical energy supplied by massive stars in
the form of supernovae and stellar winds (Leitherer \& Heckman
1995). For typical starburst parameters, the rate
of supply of mechanical energy is of-order 1\% of the bolometric
luminosity of the starburst and typically 10 to 20\% of the Lyman
continuum luminosity. 

The dynamical evolution of a starburst-driven outflow has been
extensively discussed (e.g. Chevalier \& Clegg 1985; Suchkov et al. 1994;
Wang 1995; Tenorio-Tagle \& Munzo-Tunon 1998; Strickland \& Stevens 2000).
Briefly, the deposition of mechanical energy by
supernovae and stellar winds results in an over-pressured cavity
of hot gas inside the starburst. 
The temperature of this hot gas
is given by:
\begin{displaymath}
T = 0.4 \mu m_H \dot{E}/k\dot{M} \sim 10^8 {\cal L}^{-1} K
\end{displaymath}
for a mass (kinetic energy) deposition rate of $\dot{M}$ ($\dot{E}$).
The ``mass-loading'' term $\cal L$ represents the ratio of the total mass
of gas that is heated to the mass that is directly
ejected by supernovae and stellar winds (e.g. $\cal L \geq$ 1).

This hot gas will expand, sweep
up ambient material and thus develop a bubble-like structure. 
The predicted expansion speed of the outer wall of such an adiabatic wind-blown
superbubble is of-order 10$^2$ km s$^{-1}$:
\begin{displaymath}
v_{Bubble} \sim 100 \dot{E}_{42}^{1/5} n_0^{-1/5} t_7^{-2/5} km/s
\end{displaymath}
for a bubble driven into an ambient medium with nucleon density
$n_0$ (cm$^{-3}$) by
mechanical energy deposited at a rate $\dot{E}_{42}$ (units of
10$^{42}$ erg s$^{-1}$) for a time $t_7$ (units of 10$^7$ years).

If the ambient medium is stratified (like a disk), the superbubble will
expand most rapidly in the direction of the vertical pressure
gradient. After the superbubble size reaches several disk vertical scale
heights, the expansion will accelerate, and it is believed that
Raleigh-Taylor instabilities will then lead to the fragmentation
of the bubble's outer wall (e.g. MacLow, McCray, \& Norman 1989).
This allows the hot gas to ``blow out''
of the disk and into the galactic halo in the form of a weakly
collimated bipolar outflow (i.e. the flow makes a transition from
a superbubble to a superwind).
The terminal velocity of this hot
wind is expected to be in the range of one-to-a-few thousand
km s$^{-1}$:
\begin{displaymath}
v_{wind} = (2 \dot{E}/\dot{M})^{1/2} \sim 3000 {\cal L}^{-1/2} km/s
\end{displaymath}

The wind will carry entrained interstellar material out of the galactic disk 
and into the halo, and will also interact with ambient halo clouds
(e.g. Suchkov et al. 1994; Strickland \& Stevens 2000).
This interstellar material
will be accelerated by the wind`s ram pressure
to velocities of few hundred km s$^{-1}$:
\begin{displaymath}
v_{cloud} \sim 600 \dot{p_{34}}^{1/2}\Omega_w^{-1/2}r_{0,kpc}^{-1/2}
N_{cloud,21}^{-1/2} km/s
\end{displaymath}
for a cloud with a column density $N_{cloud,21}$ (units of 10$^{21}$
cm$^{-2}$) that - starting at an initial radius of $r_0$ (kpc) - is
accelerated by a wind that carries a total momentum flux of
$\dot{p_{34}}$ (units of 10$^{34}$ dynes) into a solid angle
$\Omega_w$ (steradian).

\section{The Observational Manifestations of Superwinds}

Based on the above picture, we can broadly classify the gas
in a superwind into two categories. The first is the ambient
interstellar medium, and the second is the 
the volume-filling energetic fluid created by the thermalization
of the starburst's stellar eject. The thermal and kinetic
energy of this fluid is the ``piston'' that drives the outflow and dominates
its energy budget.
The observed manifestations
of superwinds arise when the primary wind fluid interacts hydrodynamically
with relatively dense ambient interstellar gas.

This has long been known to apply to the optical emission-line gas.
In the case of superbubbles, the limb-brightened
morphology and the classic ``Doppler ellipses'' seen in long-slit
spectroscopy of dwarf starburst galaxies
(e.g. Meurer et al. 1992; Marlowe et al. 1995; Martin 1998)
are consistent with the standard picture of emission from
the shocked outer shell
of a classic wind-blown bubble (e.g. Weaver et al. 1977). Typical
expansion velocities are 50 to 100 km s$^{-1}$. Similarly,
the morphology and kinematics of the emission-line gas in the outflows
in edge-on starbursts like M 82, NGC 253, NGC 3079, and NGC 4945
imply that this material is flowing outward on the surface of a
hollow bi-polar structure whose apices correspond to the starburst
(e.g. Heckman, Armus, \& Miley 1990; Shopbell \& Bland-Hawthorn 1998;
Cecil et al. 2001).
The deprojected outflow speeds range from a few hundred to a thousand
km s$^{-1}$.
This material is presumably ambient gas that has been entrained into
the boundary layers of the bipolar
hot wind, or perhaps the side walls of a ruptured superbubble
(e.g. Suchkov et al. 1994; Strickland
\& Stevens 2000). In both superbubbles and superwinds, the optical
emission-line gas is excited by some combination of wind-driven shocks and
photoionization by the starburst.

\begin{figure}[!t]
\plotone{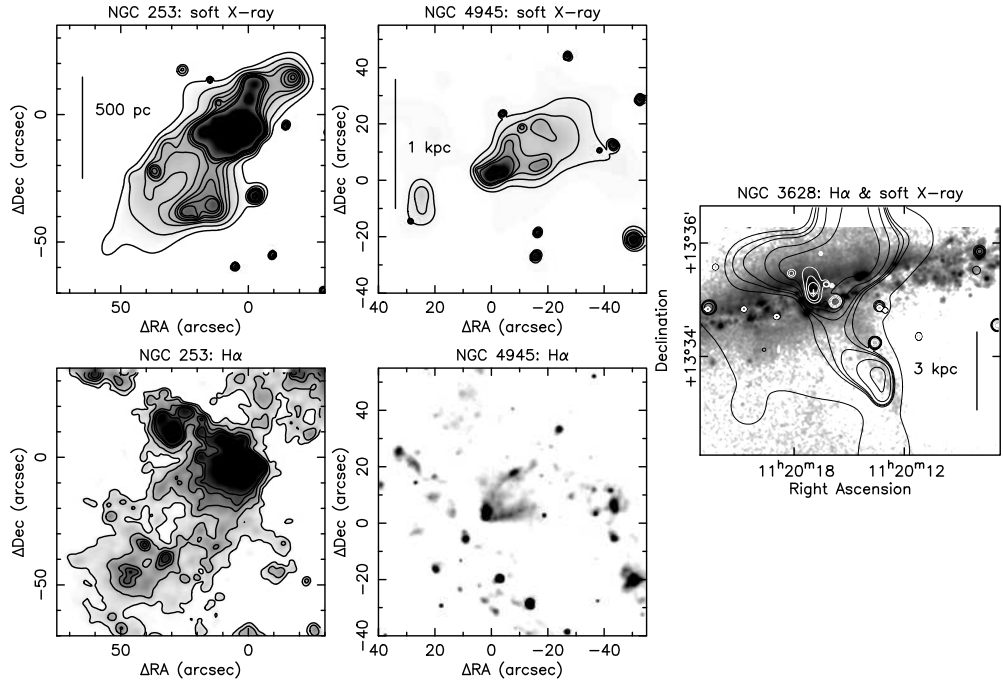}
\caption{Soft X-ray and H$\alpha$ emission in several
        edge-on starburst galaxies, showing the
        spatial similarities between the two phases.
        NGC 253 \& NGC 4945 have kpc-scale
        limb-brightened nuclear outflow cones
        (the opposite outflow cone is obscured in both cases)
        with a close match between X-ray
        \& H$\alpha$~emission. In NGC 3628 a $5$ kpc-long
        H$\alpha$ arc on the eastern limb of the wind
        is matched by an offset X-ray filament.}
\end{figure}

Prior to the deployment of the $Chandra$ X-ray observatory, it was sometimes
assumed that the soft X-ray emission associated with superbubbles
and superwinds represented the primary wind fluid that filled
the volume bounded by the emission-line gas.
If so, its relatively
low temperature (typically 0.5 to 1 kev) and high luminosity
($\sim 10^{-4}$ to $10^{-3} L_{bol}$) required that substantial
mass-loading had occurred inside the starburst ($\cal L \sim$ 10).
The situation is actually more complex.
The superb imaging capabilities of $Chandra$ 
demonstrate that the X-ray-emitting material bears a very
strong morphological relationship to the optical emission-line gas
(Strickland et al. 2000,2001; Martin, Kobulnicky, \& Heckman 2001).
The X-ray gas
also shows a limb-brightened filamentary structure,
and is either coincident with, or lies just to the ``inside'' of
the emission-line filaments (Fig. 1). Thus, the
soft X-rays could arise in regions in which hydrodynamical processes at the
interface between the wind and interstellar medium have
mixed a substantial amount of dense ambient gas into the wind fluid,
greatly increasing the local X-ray emissivity. Alternatively, the
filaments may represent the side walls of a ruptured superbubble left behind
as the wind blows-out of a ``thick-disk'' component in the
interstellar medium. In this case the H$\alpha$ emission might trace
the forward shock driven into the halo gas and the X-rays the reverse
shock in the wind fluid (see Lehnert, Heckman, \& Weaver 1999) 

Ambient interstellar material accelerated by the wind can also give
rise to blueshifted interstellar absorption-lines in starbursts.
Our (Heckman et al. 2000) survey of
the NaI$\lambda$5893 
feature in a sample of several dozen starbursts showed that 
the absorption-line profiles in the outflowing interstellar gas
spanned the range from near the galaxy
systemic velocity to a typical maximum blueshift of 400 to 600 km s$^{-1}$.
We argued this represented the terminal velocity reached by 
interstellar clouds accelerated by the wind's ram pressure.
Very similar kinematics are observed in vacuum-UV absorption-lines in
local starbursts (Heckman \& Leitherer 1997;
Kunth et al. 1998; Gonzalez-Delgado et al. 1998). This material
(Figure 2) ranges
from neutral gas probed by species like OI and CII to coronal-phase
gas probed by OVI (Heckman et al. 2001a; Martin et al. 2001).
Heckman et al. (2000) showed that there are
substantial amounts of outflowing dust associated
with the neutral phase of the superwind. Radiation pressure may play
an important role in accelerating this material (e.g. Aguirre 1999).

Extended radio-synchrotron halos around starbursts imply
that there is a magnetized relativistic component of the outflow.
In the well-studied case of M 82, this relativistic
plasma has evidently been
advected out of the starburst by the primary energy-carrying
wind fluid (Seaquist \& Odegard 1991). The situation in NGC 253
is less clear (Beck et al. 1994; Strickland et al. 2001)

\begin{figure}
\plotone{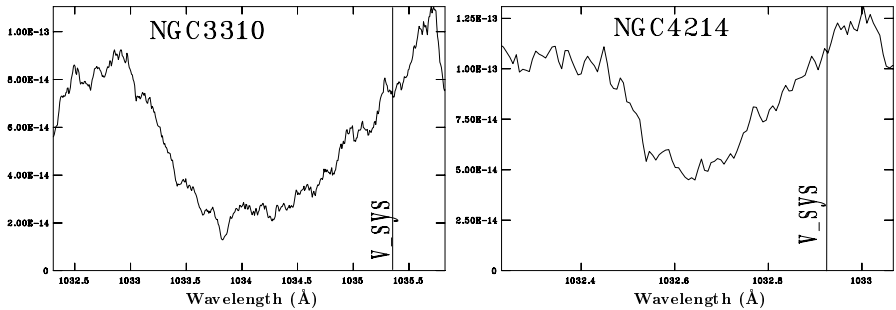}
\caption{$FUSE$ spectra of the OVI$\lambda$1031.9 interstellar absorption-line,
tracing outflowing coronal-phase gas. The absorption covers the range
from $v_{sys}$ to a maximum blueshift of $\sim$700 km s$^{-1}$ in
the powerful starburst NGC 3310 (left panel) and $\sim$140 km s$^{-1}$
in the starbursting irregular galaxy NGC 4214 (right panel).}
\end{figure}

\section{Superwind Demographics}

Lehnert \& Heckman (1996) discussed the analysis of the optical emission-line
properties of a sample of $\sim$50 disk galaxies selected to be bright and warm
in the far-infrared (active star-formers) and to be viewed within
$\sim30^{\circ}$ of edge-on.
They defined several indicators of minor-axis outflows: 1) an excess of
ionized gas along the minor axis (from H$\alpha$ images) 2) emission-line
profiles that were broader along the galaxy minor axis than along the major
axis 3) emission-line ratios that were more ``shock-like'' along
the galaxy minor axis than the major axis.  
All these indicators became stronger in the galaxies with more
intense star-formation (larger $L_{FIR}$, larger $L_{FIR}/L_{OPT}$,
and warmer dust temperatures). 

Dahlem, Weaver, \& Heckman (1998) used $ROSAT$ and $ASCA$ to search
for X-ray evidence for outflows from a complete sample of the
seven nearest edge-on starburst galaxies
(selected on the basis of far-IR flux, warm far-IR colors,
edge-on orientation, and low
Galactic HI column). Apart from the dwarf galaxy NGC 55, all the
galaxies showed hot gas in their halos. The gas had temperatures of
a few times 10$^6$ to 10$^7$ K, and could be traced out to distances
of-order 10 kpc from the disk plane 
(see also Read, Ponman, \& Strickland
1997).

Heckman et al. (2000) obtained spectra of  
19 starbursts in which the NaI$\lambda$5893 (NaD) absorption feature
was produced primarily by interstellar gas (rather than stars). In
12 of these (63\%) the NaD centroid was blueshifted by $\sim10^2$ to $10^3$
km/s relative
to the galaxy systemic velocity, and this fraction rose to 79\% in
galaxies viewed from within 60$^{\circ}$ of face-on. No comparably
redshifted absorption was seen in any galaxies.

At high-redshift, the only readily available tracers of superwinds
are the interstellar absorption-lines in the rest-frame ultraviolet.
As shown by Franx et al. (1997) and Pettini et al. (2001)
the Lyman Break galaxies generically show
interstellar absorption-lines that are blueshifted by a few hundred
to over a thousand km s$^{-1}$ relative to the estimated galaxy systemic
velocity. These galaxies strongly resemble local starbursts in their
high rate of star-formation per unit area (Meurer et al. 1997).

{\it In summary, 
superwinds are ubiquitous in galaxies with star-formation-rates
per unit area $\Sigma_* \geq$ 10$^{-1}$ M$_{\odot}$ yr$^{-1}$ kpc$^{-2}$.
Starbursts and the Lyman Break galaxies surpass
this threshold, while the disks of ordinary present-day spiral
galaxies do not (Kennicutt 1998).}

\section{Estimates of Outflow Rates}

While it is relatively straightforward to demonstrate that
a superwind is present, it is more difficult to robustly calculate the
rates at which mass, metals, and energy are being transported out by
the wind. Several different types of data can be used, each with its
own limitations and required set of assumptions.\\

{\bf X-Ray Emission:} X-ray imaging spectroscopy yields the superwind's
``emission integral''. 
Presuming that the X-ray spectra
are fit with the correct model for the hot gas it follows that
the mass and energy of the X-ray gas scale as follows:
$M_X \propto (L_X f)^{1/2}$ and
$E_X \propto (L_X f)^{1/2} T_X (1+{\cal M}^2)$. Here
$f$ is the volume-filling-factor of the X-ray gas and
$\cal M$ is its Mach number. Numerical hydrodynamical simulations
of superwinds suggest that
$\cal M$$^2$ = 2 to 3 (Strickland \& Stevens 2000).
The associated outflow rates ($\dot{M}_X$ and $\dot{E}_X$)
can then be estimated by dividing
$M_X$ and $E_X$ by the crossing time of the observed region:
$t \sim (R/c_s\cal M)$, where $c_s$ is the speed-of-sound.

If the X-ray-emitting gas is assumed to be volume-filling ($f \sim$
unity), the resulting values for $\dot{E}_X$ and $\dot{M}_X$
are then very similar
to the starburst's rates of kinetic energy deposition and
star formation respectively. As described above, $Chandra$ images 
(Fig. 1) show
that the X-ray-emitting gas does not have unit volume filling factor.
On morphological and physical grounds we have
argued that $f$ is of-order 10$^{-1}$. 
This would mean that previous estimates
of $\dot{M}_X$ and $\dot{E}_X$ are overestimated by a factor of $\sim$ 3.\\

{\bf Optical Emission:}
Optical data on the warm ($T \sim$ 10$^4$ K) ionized gas can be used
to determine the outflow rates $\dot{M}$ and $\dot{E}$ in a way that
is quite analogous to the X-ray data.
In this case, the outflow velocities
can be directly measured kinematically from spectroscopy.
Martin (1999) found the implied values for $\dot{M}$
are comparable to (and may even exceed) the star-formation rate.

In favorable cases, the densities and thermal pressures can be directly
measured in the optical emission-line clouds using 
the appropriate ratios of emission lines. The thermal pressure in these
clouds traces the ram-pressure in the faster outflowing wind that is
accelerating them (hydrodynamical simulations suggest that
$P_{ram} = \Psi P_{cloud}$, where $\Psi$ = 1 to 10). Thus, for a wind with a
mass-flux $\dot{M}$ that freely flows
at a velocity $v$ into a solid angle $\Omega$, we have
\begin{displaymath}
\dot{M} = \Psi P_{cloud}\Omega r^2/v
\end{displaymath}
\begin{displaymath}
\dot{E} = 0.5 \Psi P_{cloud}\Omega r^2 v
\end{displaymath}

Based on observations and numerical models, the values
$v \sim$ 10$^3$ km s$^{-1}$,
$\Psi \sim$ a few, and $\Omega/4\pi \sim$ a few tenths are reasonable.
The radial pressure profiles $P_{cloud}(r)$ measured in superwinds by
Heckman, Armus, \& Miley (1990) and Lehnert \& Heckman (1996) then imply
that $\dot{M}$ is comparable to the star-formation rate
and that $\dot{E}$ is comparable to the starburst
kinetic-energy injection rate (implying that radiative losses are
not severe).\\

{\bf Interstellar Absorption-Lines:}
The use of interstellar absorption-lines to determine outflows rates
offer several distinct advantages.
First, since
the gas is seen in absorption against the background starlight, there is no
possible ambiguity as to the sign (inwards or outwards) of any radial flow
that is detected, and the outflow speed can be measured directly (e.g. Fig. 2).
Second, the strength of the absorption will be related  to the
column density of the gas. In contrast, the X-ray or optical surface-brightness
of the emitting gas is proportional to the emission-measure. Thus, the
absorption-lines more fully probe the whole range of gas densities in the
outflow, rather than being strongly weighted in favor of the densest material
(which may contain relatively little mass). 

The biggest obstacle to estimating outflows rates is that the strong
absorption-lines are usually saturated, so that their equivalent width
is determined primarily by the velocity dispersion and covering factor, rather
than by the ionic column density. In the cases where the rest-UV region
can be probed with adequate signal-to-noise (Pettini et al. 2000; Heckman
\& Leitherer 1997),
the total HI column in the outflow can be measured
by fitting the damping wings of the Ly$\alpha$ interstellar line,
while ionic columns may be estimated from the weaker (less saturated)
interstellar lines. In the Heckman et al. (2000) survey of the
NaD line, we estimated NaI columns in the outflows based
on the NaD doublet ratio (Spitzer 1968), and we then
estimated the HI column assuming that the gas obeyed the same
relation between $N_{HI}$ and $N_{NaI}$ as in the Milky Way. These
HI columns agreed with columns estimated independently from the line-of-sight
color excess $E(B-V)$ toward the starburst, assuming a Galactic gas-to-dust
ratio. From both the UV data and the NaD data, the typical inferred
values for $N_{HI}$ are of-order 10$^{21}$ cm$^{-2}$.

We can then adopt
a simple model of a superwind
flowing into a solid angle $\Omega_w$ at a velocity $v$ from a
minimum radius $r_*$ (taken to be the radius of the starburst within which
the flow originates). This implies:

\begin{displaymath}
\dot{M} \sim 30 (r_*/kpc) (N_H/10^{21} cm^{-2}) (v/300 km/s)
 (\Omega_w/4\pi) M_{\odot}/yr
\end{displaymath}

\begin{displaymath}
\dot{E}  \sim 10^{42} (r_*/kpc) (N_H/10^{21} cm^{-2})
(v/300 km/s)^3  (\Omega_w/4\pi) erg/s
\end{displaymath}

Based on this simple model, Heckman et al. (2000) estimated that the
implied outflow rates of cool atomic gas are comparable to the
star-formation rates (e.g. several tens of solar masses per year
in powerful starbursts). The flux of kinetic energy carried by
this material is substantial (of-order 10$^{-1}$ of the kinetic
energy supplied by the starburst). We also estimated that $\sim$1\%
of the mass in the outflow is the form of dust grains.\\

{\bf Summary:}
The various techniques for estimating the outflow rates
in superwinds rely on simplifying assumptions (not all of which may be
warranted). On the other hand, it is gratifying that the different
techniques do seem to roughly agree: {\it the outflows carry
mass out of the starburst at a rate comparable to the star-formation
rate and kinetic/thermal energy out at a rate comparable to the rate
supplied by the starburst.}

\section{The Fate of Superwinds}

The outflow rates in superwinds should not be taken directly
as the rates at which mass, metals, and energy {\it escape} from
galaxies and are transported into the intergalactic medium.
After all, the observable manifestations of the outflow are produced by
material still relatively deep within the gravitational potential
of the galaxy's dark matter halo. We know very little about the gaseous
halos of galaxies, and it is possible that
this halo gas could confine a wind that has blown-out of a galactic
disk (Silich \& Tenorio-Tagle 2001).

A necessary condition for wind escape is that radiative losses
are not severe enough to drain energy from the wind, causing it to stall
(e.g. Wang 1995). The X-ray luminosity of the wind is typically on
of-order 1\% of the rate at which the starburst supplies kinetic energy.
Thus, radiative losses from hot ($T \geq 10^6$ K) gas will not be
dynamically significant. The radiative cooling curve peaks in the
so-called ``coronal'' regime ($T \sim 10^5$ to $10^6$ K). The $FUSE$
mission has now provided the first probe of coronal-phase gas
in starbursts and their winds via the OVI$\lambda$1032,1038 doublet
(Figure 2).
Our analysis of these data imply that
in no case is radiative cooling by the coronal gas sufficient
to quench the outflow (Heckman et al. 2001a; Martin et al. 2001).

In the absence of severe radiative cooling, one instructive
way of assessing the likely fate of the superwind material
is to compare the observed or estimated outflow velocity to
the estimated escape velocity from the galaxy.
For an isothermal gravitational potential that extends to a maximum radius
$r_{max}$, and has a circular rotation velocity $v_{rot}$, the escape velocity
at a
radius $r$ is given by:

\begin{displaymath}
v_{esc} = [2v_{rot}(1 + ln(r_{max}/r))]^{1/2}
\end{displaymath}

In the case of the interstellar absorption-lines, Heckman et al. (2000)
argued that the observed profiles were produced by material
ablated off ambient clouds and accelerated
by the wind up to a terminal
velocity represented by the most-blueshifted part of the profile.
In the case of the X-ray data, we do not measure a Doppler shift
directly, but we can define a characteristic outflow speed $v_X$
corresponding to the observed temperature $T_X$, assuming an adiabatic
wind with a mean mass per particle $\mu$ (Chevalier \& Clegg 1985):

\begin{displaymath}
v_X \sim (5kT_X/\mu)^{1/2}
\end{displaymath}

This is a conservative assumption as it ignores the kinetic energy the
X-ray-emitting gas already has (probably a factor
typically 2 to 3 times
its thermal energy - Strickland \& Stevens 2000).
Based on this approach, Heckman et al. (2000)
and Martin (1999) found that the observed
outflow speeds
are independent of the galaxy rotation speed and have typical values
of 400 to 800 km s$^{-1}$. This suggests that the outflows can readily
escape from dwarf galaxies, but possibly not from the more massive
systems.

In all these discussions it is important to keep in mind the multiphase
nature of galactic winds. It is possible (even likely) that the
question of ``escape'' will have a phase-dependent answer. The
relatively dense ambient interstellar material seen in absorption-lines,
in optical line emission, and perhaps soft X-rays
may be propelled only as far as the halo and then
return to the disk. In contrast, the primary energy-carrying wind fluid
(which could be flowing out at velocities of up to 3000 km s$^{-1}$)
could escape even the deepest galactic potentials and carry away much
of the kinetic energy and metals supplied by the starburst. Moreover,
for a realistic geometry, it is clearly much easier
for a wind to blow-out of a galaxy's interstellar medium than
than to blow it away
(e.g. De Young \& Heckman 1994; MacLow \& Ferrara 1999). 

How far out from the starburst can the effects of superwinds be observed?
In general, such tenuous material will be better traced via absorption-lines
against background QSOs than by its emission (since the emission-measure
will drop much more rapidly with radius than will the column density). To date,
the only such experiment that has been conducted is by Norman et al. (1996)
who examined two sight-lines through the halo of the merger/starburst
system NGC 520 using HST to observe the MgII$\lambda$2800 doublet. Absorption
was definitely detected towards a QSO with an impact parameter of
35 $h_{70}^{-1}$ kpc and possibly towards a second QSO with
an impact parameter of 75 $h_{70}^{-1}$ kpc. Since NGC 520 is immersed
in tidal debris (as mapped in the HI 21cm line), it is unclear
whether the MgII absorption is due to tidally-liberated or wind-ejected
gas. We can expect the situation to improve in the next few years,
as the $Galex$ mission and the $Sloan~Digital~Sky~Survey$ provide us with $10^5$
new QSOs and starburst galaxies, and the $Cosmic~Origins~Spectrograph$
significantly improves the UV spectroscopic capabilities of HST.

While a wind's X-ray surface brightness drops rapidly with radius due
to expansion and adiabatic cooling, its presence at large radii can
be inferred if it collides with an obstacle. In the case of M 82,
Lehnert, Heckman, \& Weaver (1999) show that a ridge of diffuse X-ray
and H$\alpha$
emission at a projected distance of 12 kpc from the starburst is
most likely due to a wind/cloud collision in the galaxy halo. An
even more spectacular example (Irwin et al. 1987) is the peculiar tail of
HI associated
with the galaxy NGC 3073 which points directly away from
the nucleus of its companion: the superwind galaxy NGC 3079
(50 $h_{70}^{-1}$ kpc away from NGC 3073 in projection). Irwin et al. (1987)
proposed that the HI tail is swept out of NGC 3073 by the
ram pressure of NGC 3079's superwind.

\section{Implications of Superwinds}

As discussed above, we now know that superwinds are ubiquitous in
actively-star-forming galaxies in both the local universe, and at
high-redshift. The outflows detected in the high-z Lyman Break galaxies
are particularly significant, since these objects may plausibly
represent the production sites of much of the stars and metals
in today's universe
(Steidel el al. 1999). Even if the sub-mm $SCUBA$ sources turn out
to be a distinct population at high-z, their apparent similarity
to local ``ultraluminous galaxies'' suggests that they too
will drive powerful outflows (Heckman et al. 1996,2000).
With this in mind, let me briefly describe the implications of superwinds
for the evolution of galaxies and the inter-galactic medium.

Martin (1999) and Heckman et al. (2000) showed that the estimated outflow
speeds of the neutral, warm, and hot phases in superwinds are
$\sim$ 400 to 800 km s$^{-1}$, and are independent of the rotation speed
of the ``host galaxy'' over the range $v_{rot}$ = 30 to 300 km s$^{-1}$.
This strongly suggests that the outflows selectively escape the potential
wells of the less massive galaxies. This would provide a natural
explanation for the strong mass-metallicity relation in present-day
galaxies (e.g. Lynden-Bell 1992; Tremonti et al. 2001).

As summarized above, the mass-outflow rate in entrained interstellar
matter in a superwind is similar to the star-formation rate in the
starburst.
The selective loss of gas-phase baryons from low-mass galaxies
via supernova-driven winds is an important ingredient in
semi-analytic models of galaxy formation (e.g. Somerville \& Primack 1999).
It is usually
invoked to enable the models to reproduce the observed faint-end
slope of the galaxy luminosity function by selectively suppressing
star-formation in low-mass dark-matter halos.

A different
approach is taken by
Scannapieco, Ferrara, \& Broadhurst (2000), who have argued that
starburst-driven outflows
can suppress the formation
of dwarf galaxies by ram-pressure-stripping the
gaseous baryons from out of the dark-matter halos of low-mass {\it companion}
galaxies.
The NGC 3073/3079
interaction (Irwin et al. 1987) may represent a local example.

A direct consequence of a galactic-wind origin for the mass-metallicity
relation in galactic spheroids is that a substantial fraction
of the metals today should reside in the inter-galactic medium.
This has been confirmed by X-ray spectroscopy of the intra-cluster
medium (e.g. Finoguenov, Arnaud, \& David 2001). The mean metallicity of
the present-day inter-galactic medium is not known, but the
presence
of warm/hot metal-enriched intergalactic gas
is demonstrated by the abundant population
of OVI absorption-line clouds (Tripp, Savage, \& Jenkins 2000).
If the ratio of ejected
metals to stellar
spheroid mass is the same globally as in clusters of galaxies,
then the present-day
mass-weighted metallicity of a general intergalactic medium
will be of-order $10^{-1}$ solar 
(e.g. Renzini 1997; Heckman et al. 2000). Early galactic winds
have been invoked to account for the wide-spread presence
of metals in the Ly$\alpha$ forest at high-redshift (e.g.
Madau, Ferrara, \& Rees 2001).

There is now a vigorous debate as to whether and by what means
the inter-galactic medium might have been
heated by non-gravitational sources at relatively early epochs
(e.g. Ponman, Cannon, \& Navarro 1999; Pen 1999; Tozzi \& Norman 2001;
Voit \& Bryan 2001; Croft et al. 2001). 
As a benchmark, consider the maximum amount of energy 
per inter-galactic baryon that can be supplied by galactic winds.
Star-formation with the local initial mass function (Kroupa 2001)
produces about 10$^{51}$ ergs of kinetic energy from supernovae
per 30 $M_{\odot}$ of low-mass stars ($\leq$ 1 $M_{\odot}$). The present
ratio of baryons in the intra-cluster medium to baryons in low-mass
stars is $\sim$ 6 in clusters, so the amount of kinetic energy
available in principle
to heat the intra-cluster medium is then 10$^{51}$ ergs per
180 $M_{\odot}$, or $\sim$3 keV per baryon. A similar value
would apply globally. While this upper bound is based on an assumption
of unit efficiency for the delivery of supernova energy, I have
emphasized above that the observed properties of superwinds demand
high efficiency. 

The physical state of much of the inter-galactic medium is regulated
by the meta-galactic ionizing background. QSOs
alone appear inadequate to produce the inferred background at the highest
redshifts (e.g, Madau, Haardt, \& Rees 1999). In principle, star-forming
galaxies could make a significant contribution to the background, provided
that a significant fraction of the ionizing radiation can escape the
galaxy ISM. Steidel, Pettini, \& Adelberger (2001) have 
have reported
the detection of substantial amounts of escaping ionizing radiation in
Lyman Break galaxies and have speculated that galactic superwinds
clear out channels through which this radiation
can escape. We (Heckman et al. 2001b) have considered the extant 
relevant data on present-day starbursts, and have concluded that
galactic winds may be necessary but not sufficient for creating
a globally porous interstellar medium.

Heckman et al. (2000) have summarized the evidence that starbursts are ejecting
significant quantities
of dust. {\it If}
this dust can survive a trip into the intergalactic medium and remain
intact for a Hubble time,
they estimated that the upper bound on the global amount of intergalactic
dust is $\Omega_{dust}$ $\sim$ 10$^{-4}$. While this is clearly an
upper limit, it is a cosmologically interesting one
(Aguirre 1999). Dust this abundant is probably ruled out by the
recent results by Riess et al. (2001), but intergalactic dust could 
well complicate the interpretation of the Type Ia supernova Hubble diagram.

\acknowledgements

I would like to thank my principal collaborators on the work described in this
contribution: L. Armus, D. Calzetti, M. Dahlem, R. Gonzalez-Delgado, M. Lehnert,
C. Leitherer, A. Marlowe, C. Martin, G. Meurer,
C. Norman, K. Sembach, D. Strickland, and K. Weaver.
This work has been supported
in part by grants from the NASA LTSA program and the $HST$, $ROSAT$, $ASCA$, and
$Chandra$ GO programs.



\section{References}

\end{document}